\documentclass[twocolumn]{aastex63}

\RequirePackage[normalem]{ulem} 
\RequirePackage{color}\definecolor{RED}{rgb}{1,0,0}\definecolor{BLUE}{rgb}{0,0,1} 

\usepackage{graphicx}
\usepackage{amsmath}

\newcommand{\lya}{$\mathrm{Ly}\alpha$ }

\begin{document}

\title{Diffuse \lya Halos around $\sim$300 Spectroscopically Confirmed \lya Emitters at $z\sim5.7$}

\shorttitle{LAE Halos at $z=5.7$}
\shortauthors{Wu et al.}
\revised{\today}

\author{Jin Wu}
\affiliation{Kavli Institute for Astronomy and Astrophysics, Peking University, Beijing 100871, China}
\affiliation{Department of Astronomy, School of Physics, Peking University, Beijing 100871, China}

\author[0000-0003-4176-6486]{Linhua Jiang}
\affiliation{Kavli Institute for Astronomy and Astrophysics, Peking University, Beijing 100871, China}

\author{Yuanhang Ning}
\affiliation{Kavli Institute for Astronomy and Astrophysics, Peking University, Beijing 100871, China}
\affiliation{Department of Astronomy, School of Physics, Peking University, Beijing 100871, China}


\begin{abstract}
We report the detection of diffuse \lya halos (LAHs) around star-forming galaxies at $z \sim 5.7$ by stacking 310 spectroscopically confirmed \lya emitters (LAEs). The majority of the LAEs are identified from our spectroscopic survey of galaxies at $z>5.5$. They are all located in well-studied fields with deep narrowband and broadband imaging data. We combine the LAE sample and its subsamples in the narrowband NB816 (i.e., the \lya band) and $z$ band (i.e., the continuum band). By comparing the stacked objects with their corresponding point spread functions, we clearly detect extended LAHs around these LAEs. We perform sophisticated simulations and analyses on statistical and systematic errors, and confirm that the detected halos are not caused by errors. The scale lengths of the LAHs, when described by a double-component model, range from 1.2 to 5.3 kpc with a typical value of $\sim$2 kpc. The LAH sizes from our sample are in agreement with those of individual LAEs at the similar redshift measured by VLT/MUSE, but are relatively smaller than those of photometrically selected LAEs in previous studies. We also find that LAEs with higher \lya luminosities, higher UV continuum luminosities or smaller \lya equivalent widths tend to have larger LAH sizes. Our results are consistent with a scenario that LAHs originate from the scattered light of the central galaxies by H I gas in the circumgalactic medium. 
\end{abstract}


\section{Introduction}
\label{sec:intro}

The circumgalactic medium (CGM) around galaxies encodes details to study the properties of galaxy populations ~\citep[e.g.,][]{Bahcall.Spitzer1969}. It connects galaxies and the intergalactic medium (IGM) via gas inflows (fuel for starbursts) and gas outflows (feedback) ~\citep{De09b, De09a, Veilleux.etal2005}. Thus, the spatial distribution and dynamical state of the CGM provide invaluable information about the properties, formation, and evolution of galaxies.

\lya emission is a powerful tracer of the CGM. It is intrinsically the most luminous spectral emission feature in astrophysical nebulae and thus can be observed as \lya\ emitters (LAEs) by ground-based telescopes over a broad redshift range.
Ionizing photons of young stars in star-forming galaxies ionize neutral hydrogen atoms in the interstellar medium (ISM). These photons are likely to be re-emitted as \lya photons following recombination ~\citep{Partridge.Peebles1967}. Owing to the resonant nature of the transition, these \lya photons can then be scattered by surrounding neutral hydrogen atoms in the CGM and cause the \lya emission to spread spatially. The diffuse \lya emission, or \lya halo (LAH), was predicted in high-redshift LAEs by theoretical studies with sophisticated radiative transfer models~\citep{Z11, DI12, DA12, Verhamme.etal2012, Lake15}.
Meanwhile, other physical mechanisms such as fluorescence and cooling radiation can also contribute to the spatially extended \lya emission ~\citep[e.g.,][]{Lake15, MA16}.
Comparisons between models and observations may provide crucial information to understand the physical properties of the CGM neutral gas.

Extended \lya emission has been found around nearby star-forming galaxies ~\citep[e.g.,][]{Hayes.etal2013, Hayes.etal2014} and quasars ~\citep[e.g.,][]{Rauch2008LAEHalo, Goto.etal2009}. Due to the $(1 + z)^{-4}$ cosmological surface brightness dimming effect, it is usually very difficult to detect LAHs around individual galaxies at high redshift, so the image stacking is used to push the surface brightness limit to $10^{-19} ~ \mathrm{erg} ~ \mathrm{s}^{-1}\mathrm{cm}^{-2} ~ \mathrm{arcsec}^{-2}$ or even below.

The first detections of LAHs around high-redshift galaxies were done by stacking the narrowband images of LAE and Lyman-break galaxies (LBGs) ~\citep[e.g.,][]{Hayashino2004}.  \citet{S11} found very extended LAHs ($r \approx 80$ kpc) in a sample of luminous galaxies at $2 < z < 3$ and claimed that extended \lya scattering halos were a generic property of high-redshift star-forming galaxies. They used the single exponential model to characterize the halo size.  \citet{M12} confirmed the LAH detection by co-adding images of $\sim 2100$ photometrically selected LAEs at $z\sim3.1$. They obtained a smaller LAH scale length. At higher redshifts $z = 5.7$ and 6.6, \citet{Jiang2013PPI} did not find prominent LAHs around LAEs by stacking two small samples of spectroscopically confirmed LAEs. \citet{F13} argued that the systematic errors of such stacking experiments were previously underestimated and adopted a detailed analysis of various errors in the image stacking technique. They stacked 187 LAEs at $z = 2.06$ and 241 LAEs at $z = 3.10$ separately, and detected LAHs at $z=3.1$, but did not detect LAHs at $z=2.06$. \citet{M14} used a large sample of LAEs at $z=2.2\sim6.6$ and detected LAHs at all redshifts. \citet{M16} divided 3556 LAEs at $z=2.2$ into several subsamples to investigate the relation between the LAH size and galaxy properties. \citet{X17} applied a double exponential model to analyze LAHs at $z=2.66$ and 3.78.
In these previous samples, member galaxies are mostly photometrically selected LAEs. They may include contaminants such as AGNs that can produce strong LAHs. A sample of spectroscopically confirmed galaxies is preferred for the stacking method.

In this paper, we will use 310 spectroscopically confirmed LAEs at $z \sim 5.7$ to detect LAHs. We will also investigate the dependence of the LAH size on different galaxy properties. The outline of this paper is as follows.
In \autoref{sec:data}, we describe our large LAE sample at $z \sim 5.7$ and the deep imaging data that we use for stacking.
In \autoref{sec:method}, we describe our stacking technique and one-dimentional (1D) model fitting procedure.
In \autoref{sec:error}, we perform a careful analysis of errors that could be introduced in different steps, and demonstrate that our LAH detections are robust. We present the detections of LAHs around the LAEs in \autoref{sec:result}.
In \autoref{sec:discussion}, we compare our results with previous studies and recent theoretical studies. We also discuss the constrains on the physical origin of LAHs.
Finally, we summarize our paper in \autoref{sec:summary}.

Throughout this paper, we use AB magnitudes ~\citep{Oke.Gunn1983} and adopt a cosmology parameter set of $( \Omega_m , \Omega_\lambda , H_0) = (0.3, 0.7, 70~\mathrm{km}~\mathrm{s}^{-1}~\mathrm{Mpc}^{-1})$. In this cosmology, one arcsecond corresponds to a physical size of 5.87 kpc at $z = 5.7$.

\section{Data and galaxy sample}
\label{sec:data}

We compile an initial sample that consists of 337 spectroscopically confirmed LAEs from our high-redshift galaxy survey program ~\citep{Jiang2017M2FS} and the literature ~\citep{Hu10, Shimasaku.etal2006, Kashikawa.etal2011}. For the LAEs from the literature, we still use our own images. Here we briefly introduce these imaging data. For the LAEs from our program, we further introduce the target selection criteria and the follow-up spectroscopy. More details can be found in  \citet{Jiang2013PPI} and  \citet{Jiang2017M2FS}.

\subsection{Image Data}
\label{imagedata}

Our LAEs are located in well-studied deep fields: A370, COSMOS, ECDFS (extended Chandra deep field south),  HDF, SDF (Subaru deep field), SSA17, SSA22 and SXDS (Subaru XMM-Newton deep field).
They have deep Subaru Suprime-Cam imaging data in the optical bands, especially in two narrowband filters, NB816 and NB921, which correspond to the detection of LAEs at $z \simeq 5.7$ and 6.5, respectively. The full widths half maximum (FWHM) of the two filters are about 120 and 132 \AA, respectively. In this paper, we focus on the redshift 5.7 LAEs, and use the Subaru $z'$ band ($z$ band or broadband hereafter) and the NB816 band (n band or narrowband hereafter) for  image stacking.

The images were taken with Subaru Suprime-Cam, and were retrieved from the archival server SMOKA ~\citep{Baba.etal2002}. The data were reduced, re-sampled, and co-added using a combination of the Suprime-Cam Deep Field REDuction package ~\citep[SDFRED;][]{Yagi.etal2002, Ouchi.etal2004} and our own IDL routines. After routine process for individual images, we extracted sources using SExtractor ~\citep{Bertin.Arnouts1996} and calculated astrometric and photometric solutions using SCAMP ~\citep{Bertin2006} for the final image co-addition. We re-sampled and co-added images using SWARP ~\citep{Bertin2002Swarp}. See \citet{Jiang2013PPI} and  \citet{Jiang2017M2FS} for details.

In the final combined narrowband and broadband images, we measured their surface brightness and the values range from 1.2 to $4.0 \times 10^{-18} ~ \mathrm{erg} ~ \mathrm{s}^{-1}\mathrm{cm}^{-2}\mathrm{arcsec}^{-2}$. \autoref{tab:lae_fields} shows the basic information of each field. These images will be used later in the image stacking process. We also obtained deep images in other bands, and they are only used for target selection.

\begin{table}[]
	\scriptsize
	\centering
	\begin{tabular}{ccccc}
		Field (1) & \textbf{$\mathrm{SB}_{z}$} (2) & \textbf{$\mathrm{SB}_{n}$} (3) & FWHM (4) & No. (5) \\
		\hline
		A370         &   2.54 &   3.52 & $  0.53 \pm   0.02$ & 49 \\
		A370\_new       &   3.10 &   3.97 & $  0.56 \pm   0.02$ & 8 \\
		COSMOS  &   2.19 &   1.27 & $  0.99 \pm   0.02$ & 41 \\
		ECDFS      &   2.75 &   3.71 & $  0.64 \pm   0.03$ & 8 \\
		HDF          &   2.10 &   2.11 & $  0.65 \pm   0.03$ & 16 \\
		HDF\_new   &   3.65 &   2.94 & $  0.65 \pm   0.02$ & 14 \\
		SDF         &   1.37 &   2.65 & $  0.50 \pm   0.03$ & 49 \\
		SSA17      &   6.13 &   3.36 & $  0.53 \pm   0.02$ & 8 \\
		SSA22    &   4.73 &   3.33 & $  0.71 \pm   0.03$ & 20 \\
		SSA22\_new    &   4.06 &   2.91 & $  0.50 \pm   0.02$ & 16 \\
		SXDS1      &   3.05 &   3.48 & $  0.52 \pm   0.03$ & 10 \\
		SXDS2      &   2.85 &   3.61 & $  0.55 \pm   0.02$ & 18 \\
		SXDS3      &   3.56 &   3.53 & $  0.52 \pm   0.03$ & 50 \\
		SXDS4      &   3.27 &   3.41 & $  0.54 \pm   0.02$ & 11 \\
		SXDS5      &   2.95 &   3.59 & $  0.53 \pm   0.02$ & 19 \\
		\hline
	\end{tabular}
	\caption{Information of the fields. Column (1) is the field name. Column (2)-(3) are the $1 \sigma$ surface brightness limits in the $z$ band (SB$_z$) and NB816 band (SB$_{\rm n}$), in unit of $10^{-18} ~ \mathrm{erg} ~ \mathrm{s}^{-1}\mathrm{cm}^{-2}\mathrm{arcsec}^{-2}$. Column (4) is the PSF FWHM in the NB816 band. Column (5) is the number of LAEs.}
	\label{tab:lae_fields}
\end{table}

\subsection{Target Selection and Spectroscopic Observations}
\label{targetselectionandspectrumobservation}

We are carrying out a spectroscopic survey of galaxies at $5.5<z<6.8$ in the fields listed in Table 1 \citep{Jiang2017M2FS}. We selected LAE and LBG candidates using the narrowband (or \lya) technique and the dropout technique, respectively. The selection of $z \approx 5.7$ LAE candidates was mainly based on the $i' - \mathrm{NB816}$ color. We applied color cuts to all $> 7\sigma$ detections in the NB816 band. Our selection criteria were relatively conservative compared to those used in the literature ~\citep[e.g.,][]{Taniguchi.etal2005, Kashikawa.etal2006, Ouchi.etal2008}. This allowed us to include less promising candidates and to achieve high completeness.

We use the Michigan\slash Magellan Fiber System (M2FS) on the Magellan Clay telescope ~\citep{Mateo2012M2FS} to do follow-up spectroscopic observations. M2FS is a fiber-fed, multi-object, double optical spectrograph with 256 fibers. It provides a large field-of-view (FoV) of $30'$ in diameter and high throughput in the wavelength range from 3700 to 9500 \AA. Given its large FoV and the large number of fibers, M2FS is efficient to spectroscopically identify LAEs.

We have identified 268 LAEs at $z\sim5.7$ from our spectroscopic observations, including a protocluster of galaxies at $z\sim5.7$ in the SXDS field \citep{Jiang.etal2018}. The full list of the LAEs will be published elsewhere (Y. Ning et al., in preparation). We also collected another 69 spectroscopically confirmed LAEs from the literature \citep{Hu10, Shimasaku.etal2006, Kashikawa.etal2011}. In total, we obtained an initial sample of 337 spectroscopically confirmed LAEs at $z\sim5.7$. The numbers of LAEs in different fields are shown in \autoref{tab:lae_fields}. The distributions of the absolute UV magnitudes $M_{\rm UV}$ and \lya luminosities are shown in \autoref{fig:result}.

\section{Method}
\label{sec:method}

Our main goal is to investigate the existence and properties of LAHs around $z \sim 5.7$ galaxies. The individual objects are too faint to detect their radial profiles at large radii, so we need to stack these images. Stacking analysis requires clean and homogeneous data. All our images were taken by the same telescope and instrument, and were reduced by the same pipeline. But they are in 8 different fields with different PSF, so we need to construct their PSFs and match them before image stacking. In this section, we present the technique details about the PSF matching, image stacking, and the final 1D model fitting.

\subsection{Image Masking and Sample Filtering}
\label{imagemaskingandsamplefiltering}

The images are cut into stamps with a size of $25'' \times 25''$ centerd at the positions of LAEs or point sources that are used to construct PSFs. We run SExtractor to detect all objects down to $1.5\sigma$ in each stamp, and use segmentation maps to mask out all detected pixels except for the center objects. Then, we visually inspect each object and remove a small number of images based on the following criteria:

\begin{enumerate}
	\item We remove sources within 200 pixel to the image edges to avoid the boundary effect.

	\item We remove sources that are contaminated by nearby bright objects.

	\item We remove sources with multi-peak features. This essentially removes some bright mergers or clumpy systems such as CR7 \citep{Sobral2015CR7}. These objects are intrinsically very extended and will contaminate our results.

	\item We remove sources that are likely blended with other objects.

\end{enumerate}

After applying the above selection criteria we are left with a sample of 310 clean LAEs. This sample is now  contamination free and is thus an ideal sample to study the LAH properties.

\subsection{PSF and PSF Matching}
\label{sec:psf}

Before image stacking, we homogenize the PSF of each image to the same size. This is similar to the procedure outlined by  \citet{F13}.
We run SExtractor to create the source catalog of each image and select point sources based on the compactness parameter (CLASS\_STAR $>$ 0.95) without saturation flag. We stack 100 point sources for each field in the same manner as described in the next section and measure the radial profile of each stacked image by azimuthally averaging in bins of annuli. The point sources are also visually inspected by the criteria described in the previous subsection.

\begin{figure}[tbp]
	\centering
	\includegraphics[keepaspectratio,width=0.45\textwidth,height=0.75\textheight]{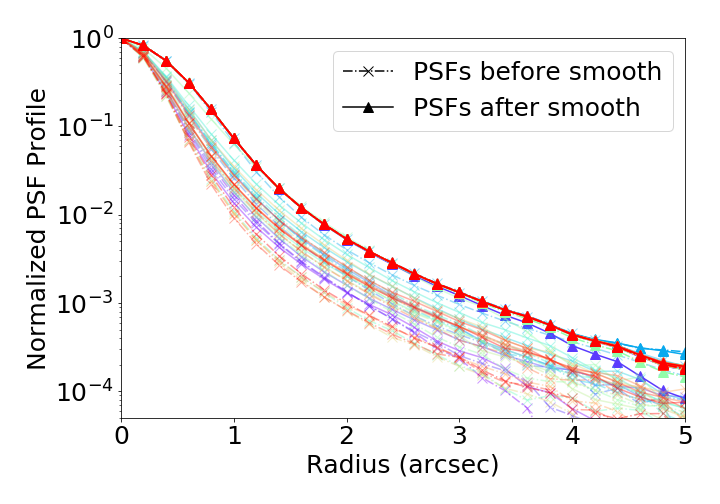}
	\caption{PSF matching results. Different colors indicate different fields. Solid and dashed lines with `x' are PSF profiles before PSF matching for the NB816 and $z$ bands, respectively. Lines with triangles are PSF profiles after PSF matching. All PSFs agree with each other after the PSF matching procedure.}
	\label{fig:psf}
\end{figure}

\autoref{tab:lae_fields} and \autoref{fig:psf} show the PSF FWHM and the PSF profile for each field. The FWHM varies from $0''.5$ in the SDF field to $0''.98$ in the COSMOS field. We use the PSF of the COSMOS field as the reference PSF since it has the largest size.
The convolution kernel is derived by comparing the PSF of each image with the reference one using the python package $\mathbf{photutils}$ \footnote{https:\slash \slash photutils.readthedocs.io}. The kernel is then used to convolve the image by the $\mathbf{numpy} ~ \mathbf{convolve}$ \footnote{http:\slash \slash www.numpy.org\slash } routine. To verify the results, we re-calculate the PSF for each field after PSF matching, and the results are shown in \autoref{fig:psf}. Now the PSFs of all fields agree with each other within $4''$.

\subsection{Image Stacking}
\label{sec:stack}

\begin{figure}[tbp]
	\centering
	\includegraphics[keepaspectratio,width=0.45\textwidth,height=0.75\textheight]{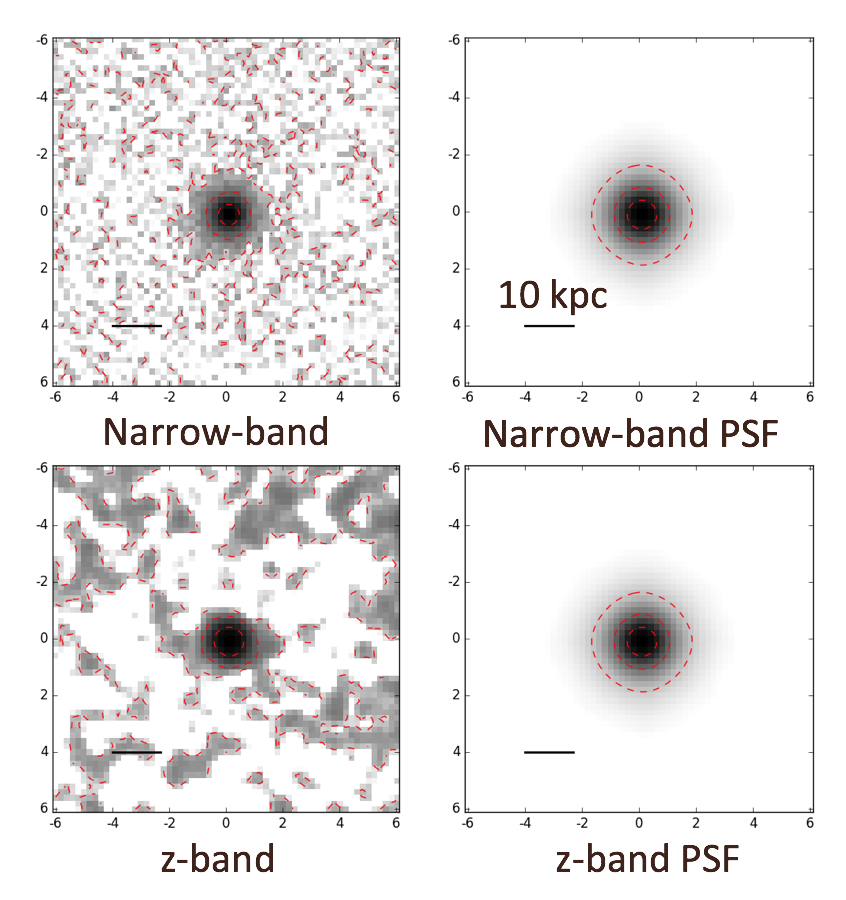}
	\caption{Stacking results for all 310 LAEs. The different panels show the stacked images in the NB816 and  $z$ bands and their corresponding PSF images. The dashed contours show the 50\%, 10\%, and 1\% flux levels, respectively.}
	\label{fig:stack}
\end{figure}

In the next step, we stack the broadband and narrowband images using SWARP. We estimate the effective local sky background using an annular region from $4''$ to $8''$ and subtract a median value from each stamp image. We adopt both the weighted-mean and weighted-median algorithms with the weight of inverse variance for each pixel. The image alignment is automatically handled by SWARP with WCS information. We then extract the 1D radial profile by azimuthally averaging each stacked narrowband image using a harsh ($2.5\sigma$) sigma-clipping algorithm and a bin step length of one pixel. Each radial bin is calculated using a full error model, including readout noise, photon noise, small-scale flat-fielding error and other systemic error (see \autoref{sec:error}). This radial profile is used for the final 1D fitting. The stacking results of all 310 LAEs are shown in \autoref{fig:stack}.

\subsubsection{Stacking Parameters}
\label{sec:techdiff}

Different parameters used in the stacking procedure may cause variations in the final results. Here we highlight some differences between our parameters and those in previous work. To investigate the LAE light profile and its central galaxy, we usually need continuum-free \lya images and line-free continuum images.
Previous studies ~\citep[e.g.,][]{F13, X17} model the LAE spectrum using a power-law model and derive these images by an image subtraction. But it is better to fit the data with a full model instead of separating the data and fit them with different model components. In this work, we do not subtract continuum flux from the \lya images. Instead, we add an extra term in the 1D fitting process. Our method can slightly improve the fitting results. Details can be found in \autoref{appendixA}. Note that our $z$-band images do not include \lya emission, so they are line-free continuum images.

Different stacking methods have been used previously, including weighted and unweighted mean and median of direct or normalized images. The median combination and radial sigma-clipping algorithm have the advantage of being robust against unwanted flux from nearby sources, but they do not preserving flux in the two-dimensional (2D) images. We adopt both weighted-mean and weighted-median algorithms. Because our sample is clean and the LAEs do not have very close objects, we will fit the 1D model using the weighted-mean profile only, and plot the weighted-median profile as a reference.

We may have included contaminant flux from projected nearby sources below our detection threshold and/or imperfect background subtraction, despite the fact that very nearby sources have been removed. The flux is indistinguishable from the true extended LAH profile at large radii. We estimate the effective local sky using an annular region from $4''$ to $8''$ and subtract its median value from the stacked image. This procedure removes the bias from the above sources but limits our ability to detect diffuse emission beyond the  region larger than $4''$. This radius is smaller than those found in previous studies ~\citep{F13, X17}. Details will be discussed in \autoref{sec:allerrors}.

\subsection{1D Model Fitting}
\label{sec:model1d}

To fit a \lya radial profile, we separate its continuum flux and get a continuum-free LAE image. We model it using a one-component exponential model (\autoref{eq:single}) and a two-component exponential model (\autoref{eq:double}). The details are shown in \autoref{appendixA}. The two models have been widely used in previous work ~\citep[e.g.,][]{S11, W16}. The fit is done by the Python MCMC package emcee\footnote{http:\slash \slash dfm.io\slash emcee}. Uncertainty of a parameter is estimated as the semi-quantile of the range enclosing the 16th and 84th percentiles of the MC distribution.

\begin{figure}[tbp]
	\centering
	\includegraphics[keepaspectratio,width=0.45\textwidth,height=0.75\textheight]{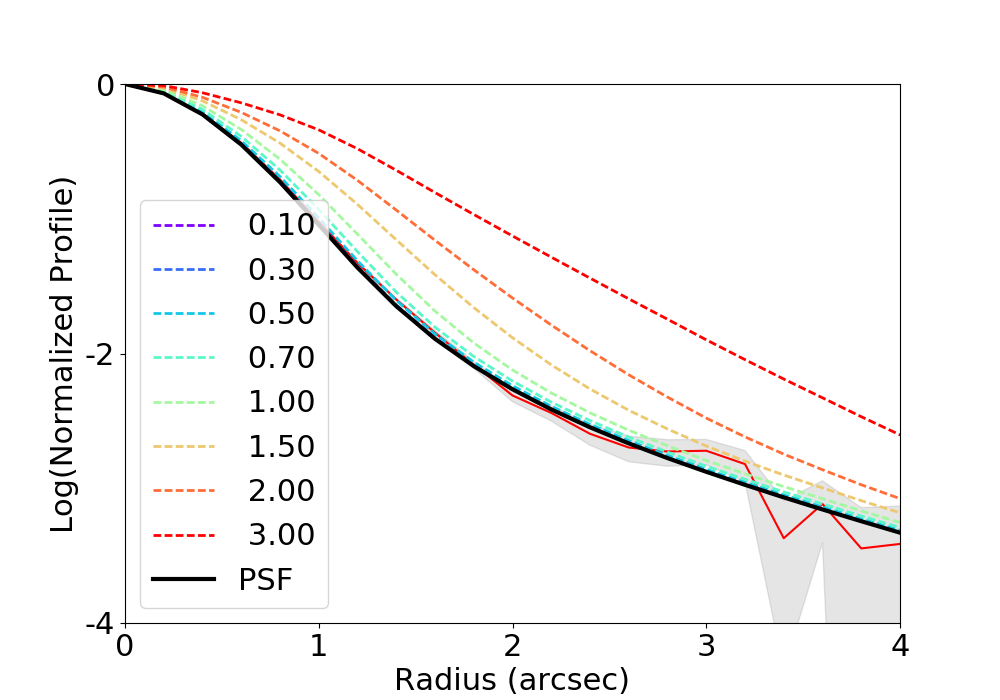}
	\caption{Stacking results of the simulated sample with positional errors. The solid red line with the shadow region is the stacked radial profile of the simulated sample with positional errors, and the shadow region shows its $1\sigma$ error region. The dashed lines with different colors show the PSF convolved with an exponential profile $\mathrm{PSF} *\mathrm{exp}(-\frac{r}{r_h})$ with different scale lengths (in unit of kpc). The solid black line is the PSF.}
	\label{fig:poserror}
\end{figure}

The one-component model uses a single exponential profile to describe the \lya and continuum profiles (\autoref{eq:single}).
This model has a linear relation between the surface brightness (in terms of mag per square arcsecs) and radius (kpc).
Using this model, the central light profile could be strongly affected by the PSF (see Appendix C of \citet{X17} for detailed discussion).
Similar to other studies, we fit the radial profile using a range of $r$ = [$1''$, $4''$]. The second model consists of two components convolved with the image PSF, a compact exponential  component $F_{g}$ and a broader halo component $F_{h}$ that declines exponentially (\autoref{eq:double}). This decomposition model automatically accounts for the PSF effect on both \lya and UV continuum images, so we can take advantage of the measured radial profile at all ranges. The examples of PSF convolved exponential profile $\mathrm{PSF} *\mathrm{exp}(-\frac{r}{r_h})$ with different scale lengths are shown in \autoref{fig:poserror}.
In these equations, $S_n$ and $S_z$ represent the narrowband and $z$-band profiles in the model, $r_g$ is the scale length of the galaxy continuum, and $r_{h,s}$ and $r_{h}$ are the scale lengths of the LAH in the single and double component models, respectively. $R^{con}_{z\Rightarrow n}$ is the factor that is used to convert the $z$-band continuum to the NB816-band flux.

\begin{eqnarray}
S_z(r)&=& S_{z0}  \exp (-\frac{r}{r_g}) \nonumber \nonumber \\
S_n(r)&=& S_{n0} \exp (-\frac{r}{r_{h,s}}) + R^{con}_{z\Rightarrow n} * S_z(r) \label{eq:single}
\end{eqnarray}

\begin{eqnarray}
S_z(r)&=& \mathrm{PSF} * S_{z} \mathrm{exp} (-\frac{r}{r_g}) \nonumber \\
S_n(r) &=& (F_{h}+ F_{g}) + R^{con}_{z\Rightarrow n} * S_z(r)\nonumber \\
&=& \mathrm{PSF} *(S_h \mathrm{exp}(-\frac{r}{r_h}) + S_g \exp (-\frac{r}{r_g}))  \label{eq:double}\\
& & + R^{con}_{z\Rightarrow n} * S_z(r)\nonumber
\end{eqnarray}

Our model is slightly different from the original version. The narrowband image contains both \lya emission flux and continuum flux. We did not subtract the continuum component in the image processing steps. Instead, we put it into the 1D model as $R^{con}_{z\Rightarrow n} * S_z(r) $. The reason is that it is better to fit the data with a full model instead of separating the data and fit them with different model components.
This factor is calculated using a power-law UV continuum model with a slope $\beta=-2$. Generally, there should also exist an extra term $R^{\mathrm{Ly}\alpha}_{n\Rightarrow z} * S_n(r)$ in the broadband that represents the \lya flux contribution in the broadband. But our NB816 filter is located outside of the $z$-band filter. The detailed discussion of this 1D model is shown in \autoref{appendixA}.

\begin{figure}[tbp]
	\centering
	\includegraphics[keepaspectratio,width=0.45\textwidth,height=0.75\textheight]{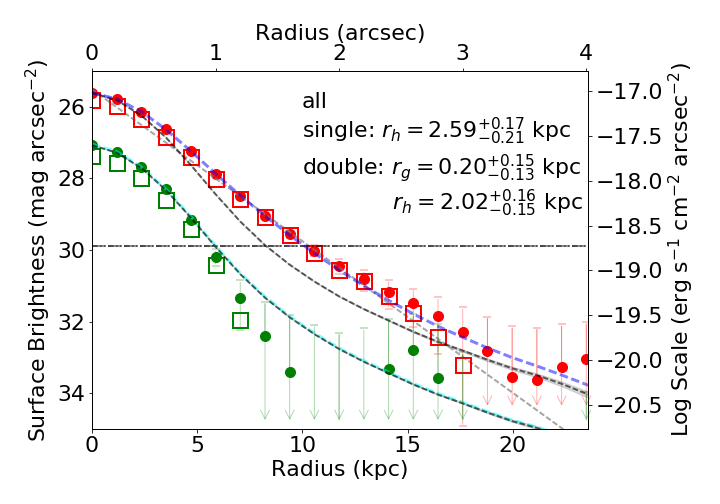}
	\caption{1D model fitting result of the full sample. The points and squares indicate the radial profile data from the average stacking and median stacking methods, respectively. The red and green colors indicate the NB816 and $z$ bands, respectively. The black dashed lines are the corresponding PSFs. The gray and blue dashed lines are the best LAH fitting results of the single and double exponential models $(S_n(r))$. The cyan dashed line is the best $z$-band fitting model $(S_z(r))$. The horizontal dashed line shows the $1\sigma$ detection limit of the stacked image (per pixel).}
	\label{fig:all}
\end{figure}

The 1D model fitting results of the full sample are plotted in \autoref{fig:all} and shown in the first row of \autoref{tab:subsamples}, where $r_{h,s}$ is the LAH size from the single exponential model and $r_h$ is the LAH size from the double exponential model. We also obtain the galaxy core scale length $r_{g}$, which is nearly zero or much smaller than $r_h$. In the next section, we will discuss the robustness of our detections. The detailed fitting results are shown in \autoref{sec:result}.

\section{Error Analyses}
\label{sec:error}

It has been argued that systematic uncertainties in stacking analyses can produce artificial extended profiles in \lya composite images ~\citep{F13, M14, M16, X17, W16}.
\citet{F13} claimed that there are two systematic sources that can produce extended \lya profile, the large-scale PSF that made by instrumental and atmospheric effects and the systematic errors of flat-fielding.  \citet{M14} further took into account the residuals of sky subtraction. In  \citet{X17} and  \citet{W16}, they also discussed the errors resulting from centroiding uncertainties. In this section, we analyze the impact of all these errors and argue that our results are not affected by them.

\subsection{Background and Systematic Errors}
\label{sec:allerrors}

The 1D model fitting results are sensitive to the sky subtraction residuals when we push the surface brightness to a very low level. In an original stamp image before stacking, we have the LAE in the center and mask the flux of other objects down to $1.5\sigma$ using the SExtractor segmentation image, so the central position still contains emission from projected nearby sources below the object detection threshold. Also, \citet{M16} found the existence of sky background oversubtraction in their UV-continuum images after several image processing steps.
To remove these biases, we estimate the effective background from an annular region at a certain distance. We choose to use an annular region $4'' \sim 8''$ from the central position to calculate the effective background. As a comparison,  \citet{X17} used an region from $6''$ to $10''$ and  \citet{F13} used an region from $13''.4$ to $18''.7$. 

Besides sky subtraction residuals, there are systematic errors from other sources (e.g. centroiding uncertainties). These errors can not be suppressed by image stacking. The inner part of a radial profile is dominated by statistic Poisson errors because of the large flux value of the object. The systematic error becomes considerably larger when the flux drops at outer regions, and will start to dominate the total error at a certain distance. The error map in our stacked image is calculated using error propagation formula and only accounts for the statistical errors. We need to add additional systematic errors at outer regions to build the full error model.

\begin{figure}[tbp]
	\centering
	\includegraphics[keepaspectratio,width=0.45\textwidth,height=0.75\textheight]{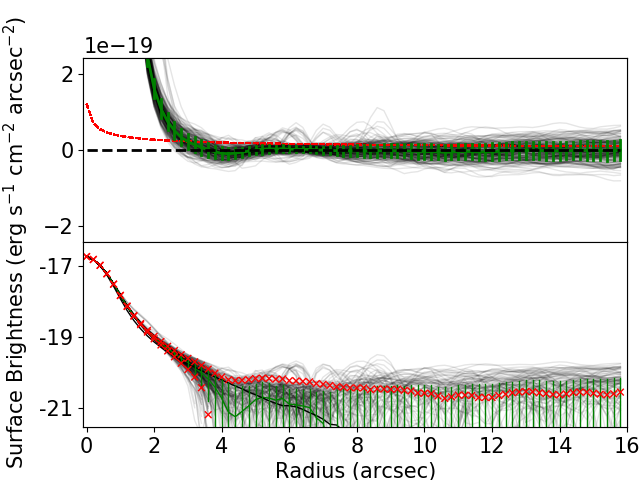}
	\caption{Effective background-corrected profiles of randomly selected objects. In the upper panel, the thin grey lines indicate effective background corrected profiles using median values in an annular region $4'' \sim 8''$. The green line with error bars shows the mean of 100 simulation profiles, which includes both statistical errors and systematic errors. The red dashed line indicates the statistical (only) errors of the green lines. The lower panel is a logarithmic version of the upper panel. The red `x' marker pairs show the regions of the average stacking results plus and minus statistical errors. The green lines with error bars are the same as those shown in the upper panel. The solid black line shows the corresponding PSF.}
	\label{fig:randomstack}
\end{figure}

We perform a series of Monte-Carlo simulations to investigate the proper region for effective background estimation and the distance where systematic errors start to dominate. We select the same number of objects from both narrowband and broadband images with similar signal-to-noise ratio (SNR) distributions. We then stack them using the same method described in \autoref{sec:stack} and extract their radial profiles. This stack of random objects is repeated 100 times. We plot these radial profiles and find that the average profile drops quickly at $<4''$ and then becomes flat with nearly zero flux at $>4''$. We thus use the median value in an annular region $4'' \sim 8''$ as the effective background. \autoref{fig:randomstack} shows the effective background-corrected profiles from our simulations.
The figure shows that the total error is almost the same as the statistical error within $8''$ and starts to rise beyond $8''$. It further shows that the average profile drops to zero near $4''$ within the $1 \sigma$ error region. Therefore, we only use data within $4''$ (20 pixel) to do 1D modeling for later analyses. We call $r_{eff} = 4''$ the effective radius because only flux within this range contain useful information and the outer part are completely dominated by background noise. We use the statistical errors as the total errors because they are almost equivalent in this region. By now we have corrected the sky subtraction residuals and completed the full error model.

We notice a small component in randomly stacked samples with halo sizes of $\lesssim 0.4$ kpc in the 1D models. This component may be caused by Image Registration Error or other unknown mechanisms. We will see that this size is much smaller than the halo size that we detect around galaxies, so it will not affect our results. In the next sections, we will discuss some mechanisms that may make our stacked images slightly more extended.

The result in \autoref{fig:randomstack} comes from a randomly selected sample without positional restrictions. If we add a positional restriction by removing nearby bright objects as described in \autoref{sec:method}, we will get a profile with positive slope in the large outer region. This is expected, because this changes the sky residual distribution from a uniform one to a biased one with more bright objects in larger radii. This phenomenon may cause a fake extended halo if we fit 1D profile using the data of outer regions. This will not affect our results since we only use data within $4''$.

\subsection{Large-scale PSF Errors}
\label{sec:psf-error}

In \autoref{sec:psf} we have built PSFs using bright but unsaturated point sources. They are small-scale PSFs as we can only get reliable profiles within $5''$. According to previous studies, the actual PSFs can be much larger, and the slopes of the PSF profiles change at large radii of $>4''$~\citep[][and references therein]{King1971, Racine1996, B2007, F13}. This could mimic the profile of an LAE in large radii and cause a fake detection of an LAH. In our work, we only use the data within $4''$, so the change of the large-scale PSF will not change our LAH profile. The only way that may affect the profile is through object residuals below the detection threshold, which is already included in our random sample simulation. We expect to get a flat background residual if objects are distributed randomly in a field. To confirm this, we model the large-scale PSF in a radius $5'' \sim 100''$ using a power-law model with different power-law indices.
We put random PSF objects with similar SNR distribution into a blank field. We then mask them using SExtractor, select random positions, and extract the 1D profiles. This is the same as we did earlier. We obtain the profiles of different background levels if we choose different PSF power-law indices and object surface densities, but all the profiles are flat. Therefore, the effect of the large-scale PSF is included in our random sample simulation.

\subsection{Image Registration Errors}
\label{imageregistrationerror}

Mis-registration of LAE centers can cause fake extended components in the final stacked images. To investigate the central offset distribution and its contribution to the halo components, we compare the extracted central positions from SExtractor with their true values from the simulation in the previous section. The target selection criterion of our LAE sample is S\slash N $>$ 7 in the narrowband and most of our targets have S\slash N $>$ 10. At SNR \ensuremath{\sim} 10, the median positional offset is $\sim 0.25$ pixel. We build simulated stamp images with one object in the center and a positional offset of $0.25$ pixels in a random direction. We choose the same number of objects as our full sample and stack them. We then extract 1D profile and perform 1D model fitting. We detect a small extend component with a scale length of $\lesssim0.45$ kpc (see \autoref{fig:poserror}). This is similar with the halo component detected in the random selected samples ($\lesssim0.4$ kpc in \autoref{sec:allerrors}) and could be the main origin of it.
This is also much smaller than the halo size ($\sim 2$ kpc) from our LAE sample. Therefore, the mis-registration of the LAE centers could contribute to the LAE profiles, but the contribution is negligible. The most flux of the extend component comes from LAH itself.

In summary, we perform Monte-Carlo simulations to determine the annular region for effective background estimation and the effects of systematic errors. We also discuss possible mechanisms to extend object profiles. We find a small extended component in the randomly selected objects which may be caused by Image Registration Error or other unknown mechanism, but its size is much smaller than the size of our LAHs, so it has negligible impact on our results.

\section{Result}
\label{sec:result}

\subsection{Diffuse LAHs around $z\sim 5.7$ LAEs}
  
The 1D model fitting results of the full sample are plotted in \autoref{fig:all}. The radial profile of the stacked image in the $z$ band is consistent with its corresponding PSF profile. The profile of the stacked image in the narrow band is clearly more extended than its corresponding PSF profile. We measure the LAH scale length $r_h$ to be $2.59^{+0.17}_{-0.21}$ kpc in the single component model and $2.02^{+0.16}_{-0.15}$ kpc in the double component model. Our sample galaxies are all spectroscopically confirmed. We have also ruled out the possibility that this apparent extended component origins from  errors. Therefore, the extended component shown in \autoref{fig:all} is from the extended LAHs around $z \sim 5.7$ LAEs.
To further confirm this detection, we perform a $\chi^2$ test. The null hypothesis is that the \lya radial profile is statistically indistinguishable from the scaled UV continuum profile when random noise is considered. We calculate $p_0$, the probability that the null hypothesis is true, and obtain $p_0<10^{-5}$ for the full sample, meaning a statistical difference between the \lya profile and UV continuum profile.
Moreover, we calculate the ratio of the total \lya model flux to the flux within its $1\arcsec$ radius ($F/F_{1\arcsec}$). This is another parameter to characterize LAHs. For the full sample, we get $F/F_{1\arcsec} = 1.28$, which means the total \lya flux is 1.28 times the flux measured within a $2\arcsec$ aperture in diameter.

\subsection{Subsamples}
\label{sec:subsample}

\begin{figure*}
	\includegraphics[keepaspectratio,height=1\textheight]{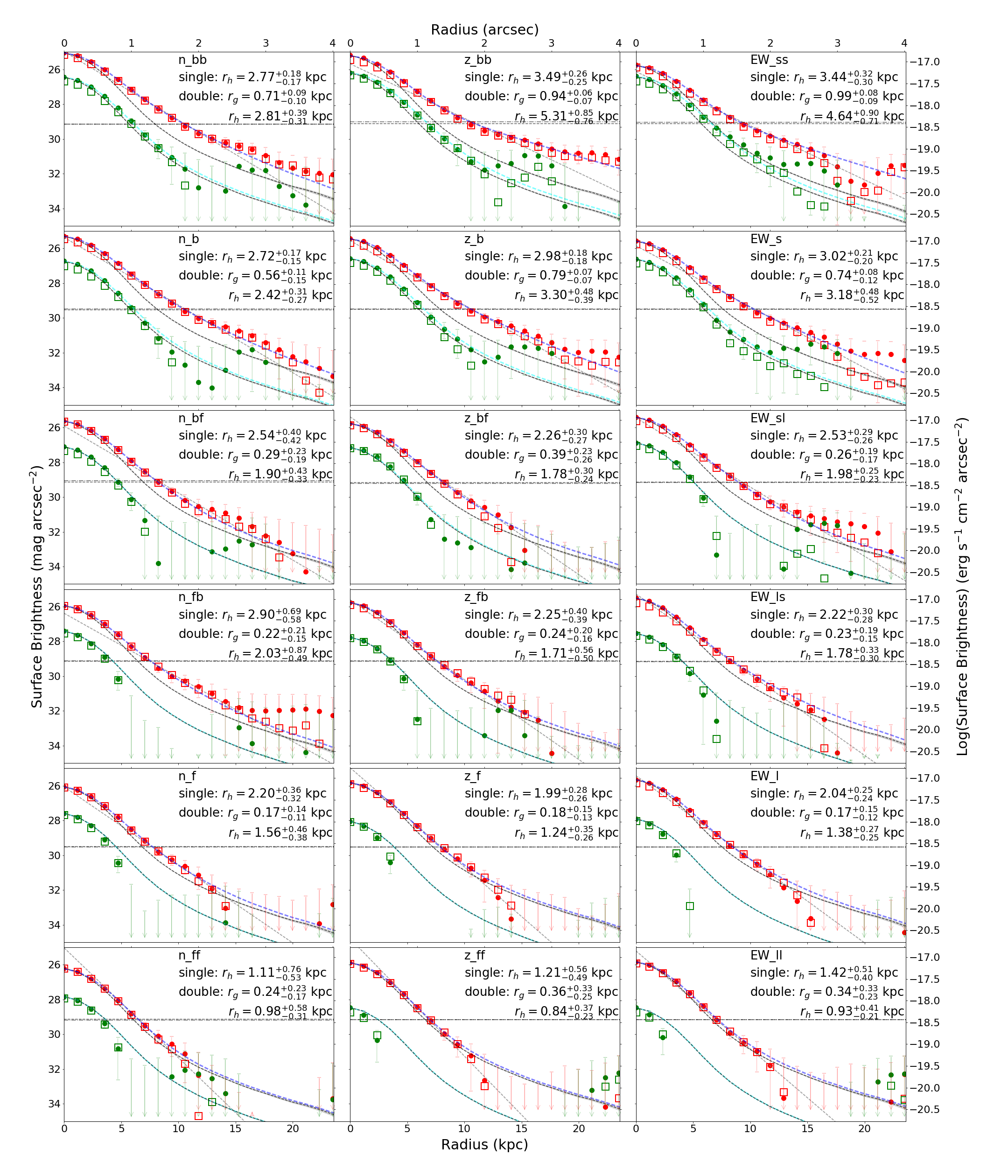}
	\caption{1D profile fitting results of all subsamples. See \autoref{fig:all} for the detailed descriptions of all plots and markers. In this figure, we find that LAEs with smaller EWs, larger narrowband flux, or larger $z$-band flux tend to have larger LAHs.}
	\label{fig:subsample_profile}
\end{figure*}

To study possible correlations between LAH profiles and LAE properties, we divide our LAE sample into several subsamples based on the absolute UV magnitude $M_{\rm UV}$, \lya luminosity, and \lya  rest-frame EW. The EW is estimated from the narrowband--broadband color. The full sample is divided by the median values of these measurements. 
The `all' sample is the full sample. The `n\_f' (faint in the narrowband) and `n\_b' (bright in the narrowband) subsamples come from the bright half and faint half of the full sample. The `z\_f' (faint in the $z$ band), `z\_b' (bright in the $z$ band), `EW\_l' (large \lya EW), and `EW\_s' (small \lya EW) subsamples are defined in the same way.
For the subsamples, we further divide them by their median values. For example, `EW\_ll' contains objects with large \lya EW in the `EW\_l' subsample, and `z\_bf' includes objects with large UV magnitude (faint flux) in the `z\_b' subsample. We put subsamples defined by the same property into one group. The selection criteria and properties of the subsamples are listed in \autoref{tab:subsamples}.
In the end, we have 19 subsamples and 4 groups. The smallest subsample contains more than 77 objects and can still provide a robust stacking result. The luminosity distributions of these subsamples and their fitting results are shown in \autoref{fig:subsample_profile} and \autoref{fig:result}. In \autoref{fig:result}, we have almost the same median UV magnitude in the continuum faint subsamples `EW\_l', `EW\_ll', `z\_f', and `z\_ff'. This is because our narrowband-selected LAEs are usually very faint in the $z$ band. Nearly half of them are below our $3\sigma$ detection limit in the $z$ band. All LAEs in the `EW\_ll' and `z\_ff' subsamples are undetected in $z$.

\subsection{Halo Sizes and Galaxy Properties}
\label{halosizeandgalaxiesproperties}

Our full sample contains 310 LAEs.  We find a scale length range of $1.1 \sim 3.5$ kpc using the single exponential model and $0.9 \sim5.3$ kpc using the double exponential model. This size range is in good agreement with the VLT MUSE measurement results of individual LAEs at similar redshifts ~\citep{W16, L17}. Most of our LAE subsamples have compact galaxy core components with nearly zero $r_g$. For those with non-negligible core components, their core scale lengths are much smaller than those of the LAH components. The results are shown in \autoref{tab:subsamples}.

In previous studies that have used the single exponential model, they reported larger LAH scales around 6$\sim$24.6 kpc, and the surface brightness profiles show extended components at large radii within their effective radii. We have a relatively small effective radius, which is more easily impacted by the center PSF. So it is not sufficient to detect LAHs using the single exponential model. In this work, we report the detection of LAHs using the double component model and the $\chi^2$ test. We claim a reliable detection of an LAH if $p_0<10^{-5}$. Based on this criterion, we report existence of LAHs for all our subsamples except for the three faintest subsample `z\_ff', `n\_ff', and `EW\_ll'.

\begin{figure}[htbp]
	\centering
	\includegraphics[keepaspectratio,height=0.3\textheight]{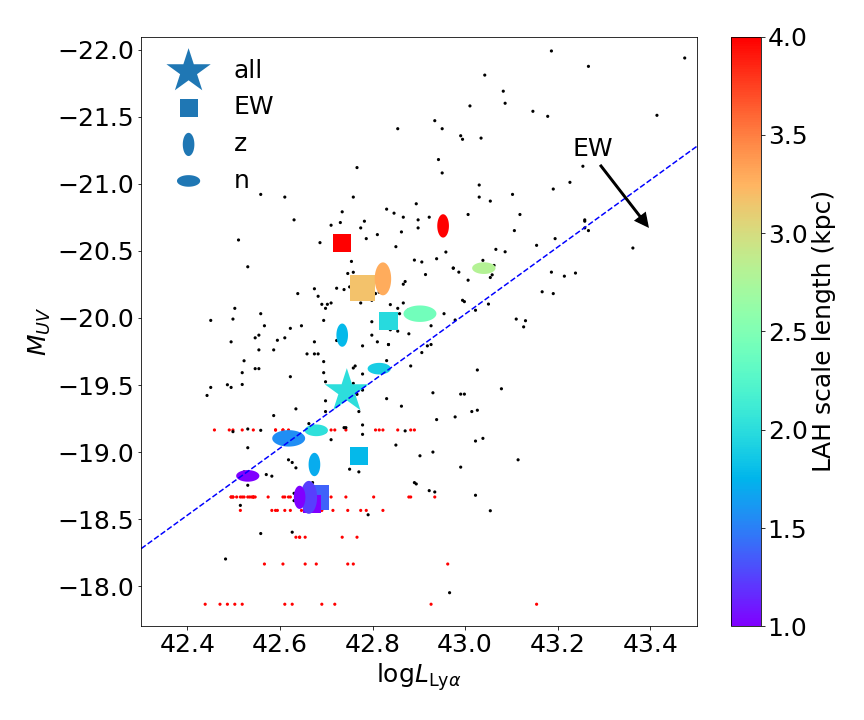}
	\caption{LAH scale lengths from the double exponential model for all subsamples. Small black and red points represent the 310 LAEs. The red points are the objects undetected in $z$, and their $y$ values are the $1 \sigma$ detection limits). The blue dashed line represents the median EW of our sample: smaller EWs are above this line and larger EWs are below this line. Color-coded symbols indicate our scale length fitting results of all subsamples in different groups. Their $x$ and $y$-axis values are the median NB816 and $z$ band luminosities of each subsample, and their colors show the halo scale sizes. The symbol sizes represent relative sample sizes. See \autoref{sec:subsample} for detailed properties of all subsamples. LAEs with smaller \lya EWs, larger narrowband flux, and/or larger $z$-band flux apparent to have larger LAE halos.
	}
	\label{fig:result}
\end{figure}

\begin{figure*}
	\includegraphics[keepaspectratio,height=0.3\textheight]{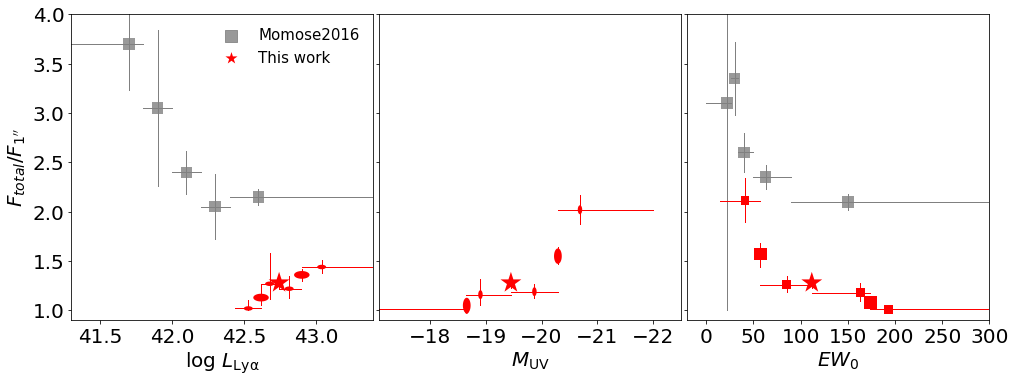}
	\caption{Ratio of total \lya model flux over the flux within its $1\arcsec$ radius. For data points of \citet{M16}, the total flux is estimated by flux within $40$ kpc.
		The red markers are defined in \autoref{fig:result}. }
	\label{fig:flux_ratio}
\end{figure*}

\begin{figure*}
	\includegraphics[keepaspectratio,height=0.50\textheight]{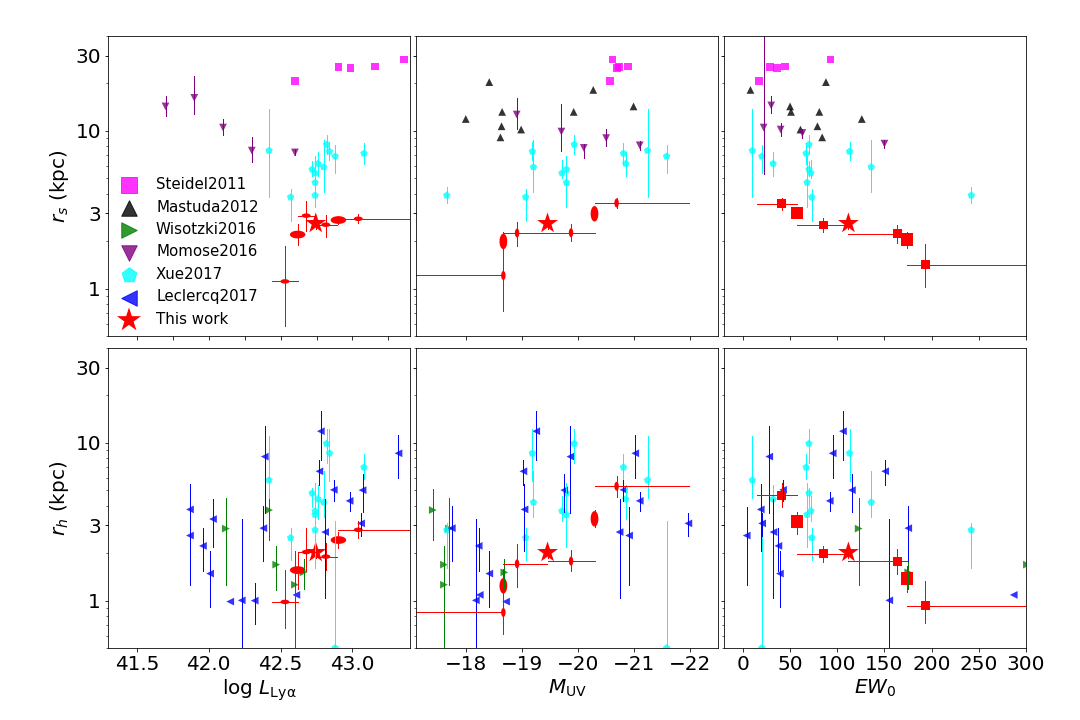}
	\caption{LAH scale lengths as a function of \lya luminosity, absolute UV magnitude, and \lya EW. The upper panels show the results from the single exponential fitting and the lower panels show the results from the double exponential fitting. Our results are shown in red, and different symbols represent different groups (see the definition in
		\autoref{fig:result} and \autoref{tab:subsamples}). In each figure, we only plot the x-axis error bar for the four smallest subsamples.  We also show scale lengths from \citet{S11}, \citet{M12}, \citet{M16}, \citet{X17}, \citet{W16}, and \citet{L17}. For the results of individual LAEs in \citet{W16} and \citet{L17}, we only plot their $z>5$ samples.
	}
	\label{fig:compare}
\end{figure*}

Using the LAH scale length measurements, we find that LAEs with smaller EWs, larger narrowband flux, or larger $z$-band flux tend to have larger LAHs. \autoref{fig:subsample_profile} shows the median luminosity distribution of all subsamples as well as their double exponential halo scale lengths. \autoref{fig:result} shows the profiles of all subsamples. In this figure, different symbols represent different groups (described in \autoref{tab:subsamples} and \autoref{sec:subsample}). Subsamples in the same group do not overlap. For each pair subsamples in the same parent sample, the one with lower broadband or narrowband luminosities has a smaller halo. The faint, large EW subsamples located in the lower left corner have very small halo scale length, while the bright, small EW subsamples in the upper right have much larger halo sizes.
\autoref{fig:flux_ratio} shows $F/F_{1\arcsec}$ for different subsamples. We get a similar trend as the scale length measurements.
We compare our results with previous studies in \autoref{fig:compare}. \citet{X17} reported similar positive correlation between LAH size and luminosity.  \citet{M16} found the same size trend with $M_{\mathrm{UV}}$ and EW, but an opposite trend with \lya luminosity. \citet{W16, L17} did not found apparent correlation in their scale length measurements.

There are several explanations for the relations that we found. LAH properties partly depend on the spatial distribution and dynamical state of the CGM. UV bright galaxies in larger dark matter halos likely have more nearby atomic hydrogen gas to scatter \lya photos. 
In simulations, \citet{DA12} found that \lya flux in narrow bands is seriously affected by the inhomogeneous IGM distribution for individual galaxies. This inhomogeneity will make the stacked profile more extend if the sample LAEs have higher \lya luminosities and a random inclination angle distribution.
The EW of an LAE is measured from the narrowband--broadband color, and thus correlates with these two parameters.  For a given \lya luminosity, a larger \lya EW means less \lya photos scattered into CGM and thus less contribution to the halo component. This is a simple explanation for the anti-correlation between the LAH size and \lya EW.

\begin{deluxetable*}{cccccccccccc}
	
	\tablecaption{All subsamples and their fitting result \label{tab:subsamples}} 
	\tablewidth{0pt}
	\tablehead{
		\colhead{Name} & \colhead{Selection} & \colhead{group} & \colhead{N} &
		\colhead{$\log{L_{Ly\alpha}}$} & \colhead{$M_{UV}$} & \colhead{EW} &
		\colhead{$r_{h,s}$} & \colhead{$r_{g}$} & \colhead{$r_{h}$} & \colhead{$F/F_{1"}$} & \colhead{$p_0$}
	}
	\decimalcolnumbers
	\startdata
	all & all objects & all & 310 & 42.74 & -19.45 & 111.8 & $2.59^{+0.17}_{-0.21}$ & $0.2^{+0.15}_{-0.13}$ & $2.02^{+0.16}_{-0.15}$ & $ 1.28^{+0.05}_{-0.05}$ & $<10^{-5}$\\
	n\_b & n $\leq$25.2 in all sample & n & 155 & 42.90 & -20.03 & 106.9 & $2.72^{+0.17}_{-0.15}$ & $0.56^{+0.11}_{-0.15}$ & $2.42^{+0.31}_{-0.27}$ & $ 1.36^{+0.06}_{-0.06}$ & $<10^{-5}$\\
	n\_f & n $>$25.2 in all sample & n & 155 & 42.62 & -19.10 & 122.5 & $2.20^{+0.36}_{-0.32}$ & $0.17^{+0.14}_{-0.11}$ & $1.56^{+0.46}_{-0.38}$ & $ 1.13^{+0.14}_{-0.08}$ & $<10^{-5}$\\
	z\_b & z $\leq$27.2 in all sample & z & 155 & 42.82 & -20.29 & 57.3 & $2.98^{+0.18}_{-0.18}$ & $0.79^{+0.07}_{-0.07}$ & $3.30^{+0.48}_{-0.39}$ & $ 1.55^{+0.09}_{-0.08}$ & $<10^{-5}$\\
	z\_f & z $>$27.2 in all sample & z & 155 & 42.66 & -18.66 & 171.8 & $1.99^{+0.28}_{-0.26}$ & $0.18^{+0.15}_{-0.13}$ & $1.24^{+0.35}_{-0.26}$ & $ 1.05^{+0.07}_{-0.03}$ & $<10^{-5}$\\
	EW\_l & EW$>$111.8$\AA$ in all sample & EW & 155 & 42.68 & -18.66 & 173.8 & $2.04^{+0.25}_{-0.24}$ & $0.17^{+0.15}_{-0.12}$ & $1.38^{+0.27}_{-0.25}$ & $ 1.08^{+0.07}_{-0.04}$ & $<10^{-5}$\\
	EW\_s & EW$\leq$111.8$\AA$ in all sample & EW & 155 & 42.78 & -20.22 & 57.2 & $3.02^{+0.21}_{-0.20}$ & $0.74^{+0.08}_{-0.12}$ & $3.18^{+0.48}_{-0.52}$ & $ 1.57^{+0.11}_{-0.13}$ & $<10^{-5}$\\
	n\_bb & n $\leq$24.8 in n\_b & n & 78 & 43.04 & -20.37 & 105.5 & $2.77^{+0.18}_{-0.17}$ & $0.71^{+0.09}_{-0.10}$ & $2.81^{+0.39}_{-0.31}$ & $ 1.44^{+0.07}_{-0.06}$ & $<10^{-5}$\\
	n\_bf & n $>$24.8 in n\_b & n & 77 & 42.81 & -19.62 & 110.9 & $2.54^{+0.40}_{-0.42}$ & $0.29^{+0.23}_{-0.19}$ & $1.90^{+0.43}_{-0.33}$ & $ 1.22^{+0.13}_{-0.09}$ & $<10^{-5}$\\
	n\_fb & n $\leq$25.5 in n\_f & n & 78 & 42.68 & -19.16 & 114.3 & $2.90^{+0.69}_{-0.58}$ & $0.22^{+0.21}_{-0.15}$ & $2.03^{+0.87}_{-0.49}$ & $ 1.27^{+0.31}_{-0.15}$ & $<10^{-5}$\\
	n\_ff & n $>$25.5 in n\_f & n & 77 & 42.53 & -18.82 & 123.6 & $1.11^{+0.76}_{-0.53}$ & $0.24^{+0.23}_{-0.17}$ & $0.98^{+0.58}_{-0.31}$ & $ 1.02^{+0.09}_{-0.02}$ & $0.17919$\\
	z\_bb & z $\leq$26.3 in z\_b & z & 78 & 42.95 & -20.69 & 45.9 & $3.49^{+0.26}_{-0.25}$ & $0.94^{+0.06}_{-0.07}$ & $5.31^{+0.85}_{-0.76}$ & $ 2.02^{+0.15}_{-0.14}$ & $<10^{-5}$\\
	z\_bf & z $>$26.3 in z\_b & z & 77 & 42.73 & -19.87 & 72.0 & $2.26^{+0.30}_{-0.27}$ & $0.39^{+0.23}_{-0.26}$ & $1.78^{+0.30}_{-0.24}$ & $ 1.19^{+0.08}_{-0.06}$ & $<10^{-5}$\\
	z\_fb & z $\leq$27.9 in z\_f & z & 78 & 42.67 & -18.91 & 157.3 & $2.25^{+0.40}_{-0.39}$ & $0.24^{+0.20}_{-0.16}$ & $1.71^{+0.56}_{-0.50}$ & $ 1.16^{+0.16}_{-0.11}$ & $<10^{-5}$\\
	z\_ff & z $>$27.9 in z\_f & z & 77 & 42.64 & -18.66 & 190.6 & $1.21^{+0.56}_{-0.49}$ & $0.36^{+0.33}_{-0.25}$ & $0.84^{+0.37}_{-0.23}$ & $ 1.01^{+0.03}_{-0.01}$ & $0.16407$\\
	EW\_ss & EW$\leq$ 57.2 $\AA$ in EW\_s sample & EW & 77 & 42.73 & -20.56 & 41.0 & $3.44^{+0.32}_{-0.30}$ & $0.99^{+0.08}_{-0.09}$ & $4.64^{+0.90}_{-0.71}$ & $ 2.11^{+0.23}_{-0.21}$ & $<10^{-5}$\\
	EW\_sl & EW$>$57.2$\AA$ in EW\_s sample & EW & 78 & 42.83 & -19.98 & 85.3 & $2.53^{+0.29}_{-0.26}$ & $0.26^{+0.19}_{-0.17}$ & $1.98^{+0.25}_{-0.23}$ & $ 1.26^{+0.09}_{-0.07}$ & $<10^{-5}$\\
	EW\_ls & EW$\leq$ 173.8 $\AA$ in EW\_l sample & EW & 77 & 42.77 & -18.97 & 163.3 & $2.22^{+0.30}_{-0.28}$ & $0.23^{+0.19}_{-0.15}$ & $1.78^{+0.33}_{-0.30}$ & $ 1.18^{+0.10}_{-0.08}$ & $<10^{-5}$\\
	EW\_ll & EW$>$173.8$\AA$ in EW\_l sample & EW & 78 & 42.67 & -18.61 & 193.2 & $1.42^{+0.51}_{-0.40}$ & $0.34^{+0.33}_{-0.23}$ & $0.93^{+0.41}_{-0.21}$ & $ 1.01^{+0.04}_{-0.01}$ & $0.00347$
	\enddata
	
	\tablecomments{ Collumn contents are:\\
		(1) Name of this subsample. \_b and \_f means bright and faint. \_l and \_s means equivalence width large and small.\\
		For example, n\_b means bright subsamples in all sample, n\_bf means faint subsamples in n\_b subsample.\\
		(2) Subsample selection criterion.\\
		(3) The group name of this subsample. \\
		(4) Number of LAEs in this subsample. \\
		(5) Median $\log {L_{Ly\alpha}}$ of this subsample. \\
		(6) Median $M_{UV}$ of this subsample. \\
		(7) Median equivalence width of this subsample. \\
		(8) Halo size of extend LAH $r_{h,s}$ in single exponential model.  \\
		(9) Halo size of galaxy core $r_{g}$ in double exponential model.  \\
		(10) Halo size of extend LAH $r_{h}$ in double exponential model.  \\
		(11) The ratio of total flux and flux with in $1"$ \\
		(12) The probability that null hypothesis is true. The null hypothesis is that the \lya radial profile is statistically indistinguishable from the scaled UV continuum profile plus random noise. \\		
	}
	
\end{deluxetable*}


\section{Discussion}
\label{sec:discussion}

\subsection{Comparison with Previous Observations}
\label{comparewithpreviousobservations}

In previous studies of LAHs at similar redshift range,  \citet{M14} stacked a sample of 397 photometrically selected LAEs at $z \sim 5.7$ and obtained an LAH scale length of $r_{h,s}=5.9^{+0.65}_{-0.53}$ for the single exponential model. This size is larger than the LAH sizes in our results. \citet{W16} and  \citet{L17} measured the scale lengths of 24 individual LAEs at $z > 5$  in the MUSE deep field. The median $r_h$ value of their measurements  is 2.86 kpc for the double exponential model, which is consistent with our results.

In previous studies of LAHs at lower redshifts around $2\sim3$, large LAH sizes have been detected using the stacking technique. For example, \citet{S11} obtained a large halo size of $r_{h,s}=24.6$ kpc in their LAE-only sample. Most of the later studies were based on photometrically selected LAE samples. \citet{M12} detected $r_{h,s}=9.1 \sim 20.4$ kpc in their subsamples.  \citet{F13} found a scale length range of $2.8 \sim 6.0$ kpc in their $z \sim 3$ sample. \citet{M14} and \citet{M16} obtained a size range of $7.3 \sim 16.2$ kpc.  \citet{X17} used both models, and their results were $r_{h,s}=5.0 \sim 7.1$ kpc and $r_h=3.9 \sim 8.6$ kpc.

For the $F/F_{1''}$ measurements (\autoref{fig:flux_ratio}), we obtain a range of $1.0 \sim 2.1$ while \citet{M16} have a much larger range of $2\sim4$. This is because of our smaller scale lengths. We obtain a similar anti-correlation between this ratio and EW. Our sample covers a small \lya luminosity range, and it is located in the same trend of $F/F_{1''}$ vs. \lya luminosity (left panel of Figure 9). In  previous studies, LAE aperture photometry was usually done within a diameter size of $2''\sim3''$. If LAHs origin from the central galaxies, their photometry was underestimated.

We discuss possible reasons that the LAH sizes in our sample are smaller than those from some previous studies. 
The first reason is that different samples are at different redshifts. The sizes of galaxies are observed to be smaller at higher redshifts \citep{Ferguson.etal2004,Shibuya.etal2019}. At high redshift, the typical half-light radii of star-forming galaxies decrease from several kpc at $z\sim3$ to $ \lesssim 1$ kpc at $z\sim6$ \citep{Jiang2013PPII}. Smaller galaxies naturally produce smaller LAHs. But this is not likely a dominated mechanism. \citet{M14} and \citet{L17} have shown that there is no obvious redshift evolution of LAHs. The second reason is that different samples may have different properties.  
Our LAEs are narrowband-selected galaxies. They have relatively high \lya luminosities and low continuum luminosities compared to those in some previous studies. 
\citet{S11} and \citet{B16} found that broadband-selected LBGs have larger LAHs than LAEs do. A similar trend is also found in our sample, i.e., lower-EW LAEs tend to have larger LAHs. However, our sample is dominated by LAEs with high \lya\ EWs. Therefore, our LAE sample has smaller LAH sizes than those in the previous LBG samples \citep{S11, M12}.  


A major difference between our sample and previous samples is that our objects are all spectroscopically confirmed. To investigate the influence of non-LAEs, we apply a similar analysis to our LAE candidate sample \citep{Jiang2017M2FS}. This sample contains 499 LAE candidates and all of them were observed by our M2FS project. Among them, 268 objects are spectroscopically confirmed LAEs and the remaining are non-LAEs. With our visual inspection (see details in \autoref{imagemaskingandsamplefiltering}), 21 objects are apparently extended or contaminated by nearby sources. We divide the sample to four subsamples: `all' sample with 499 candidates, `all-clean' sample with 478 objects, `LAE' sample with 268 LAEs, and `LAE-clean' sample with 247 clean LAEs. They have very similar median magnitudes in the NB and $z$ bands. The double component scale lengths are $1.86^{+0.12}_{-0.12}$, $1.95^{+0.13}_{-0.12}$, $2.15^{+0.20}_{-0.17}$, and $2.36^{+0.23}_{-0.19}$ kpc in the `all-clean', `all', `LAE-clean', and `LAE' subsamples, respectively. 
We can see that the `non-clean' subsamples have larger halo sizes (`all' vs. `all-clean' and `LAE' vs. `LAE-clean'), because they include a small fraction of extended objects or objects contaminated by nearby sources. On the other hand, the `all' subsamples have smaller halo sizes, because they include a significant fraction of random and compact objects (e.g., low-redshift galaxies or faint stars) that should not have diffuse halos (see also \autoref{sec:allerrors}). Therefore, the overall effect of contaminants in a sample of LAE candidates is complicated, depending on the types and relative numbers of the contaminants.




\subsection{Comparison with Theoretical Studies and the Origin of LAHs}
\label{comparewiththeoreticalstudiesandtheoriginoflah}

The existence of LAHs around star-forming galaxies in the redshift range $z=2 \sim 6$ has been confirmed, but the physical origin of LAHs is still under debate. Theoretical studies have proposed several physical origins, including the scattered light of the central galaxies by H I gas in the CGM, the contribution from satellite galaxies, and the gravitational cooling radiation from infalling gas.

One natural explanation is the scattered light scenario as \lya photons are produced by H II regions in star-forming galaxies.  \citet{Z11} performed detailed radiative transfer calculations to simulate the scattering process of \lya photons for $z\sim 5.7$ galaxies in the SXDS field. Note that about 1/3 of the LAEs in our sample are from this field. They constructed narrowband (NB816) \lya images from their simulation results and stacked these images. They also extracted  a 1D surface brightness profile that can be compared directly with observations. In this work, \lya photons are initially emitted from the center galaxies and then scattered into the CGM, contributing all the flux of LAHs. They found that the radial profile has two characteristic scales associated with two rapid changes in the profile slope, an inner scale of $4''$ and an outer scale of $40''$. The radial profile drops rapidly from the center towards the inner characteristic radius $4''$, then enters a very low  surface brightness plateau in the range of $4''-40''$. They explained that the inner profile within $4''$ is caused by the central source (one-halo term) while the outer part includes contributions from neighboring clustered sources (two-halo term). As described in \autoref{sec:allerrors}, we also detect a rapid decrease of the radial profile within $4''$. This is well consistent with \citet{Z11}, suggesting that the halo component that we detected are mostly from the one-halo term, i.e., scattered light from the central source.

The second scenario is that LAHs are produced by satellite galaxies.  \citet{Lake15} considered all potential sources of LAHs, including star formation from central galaxies, satellite galaxies, \lya photos associated with tidally stripped materials in the halo, and the gravitational cooling radiation. They concluded that a significant fraction of their simulated LAHs came from off-center star formation or gravitational cooling. In our work, we have removed objects with multi-peak features. We have also stacked $z$-band (rest-frame UV continuum) images, and the radial profiles in the $z$ band are consistent with PSF. If the LAHs that we detected were produced by satellite galaxies, we would have detected extended halos in the $z$ band. Therefore, the LAHs that we detected within $4''$ are unlikely produced by satellite galaxies.

The third scenario is that the LAH flux comes from the gravitational cooling radiation of infalling gas. Cosmological hydrodynamical simulations suggest that dense and cold gas ($\sim 10^4$ K) inflows could cause intense star formation in high-redshift galaxies ~\citep{De09b, De09a}. These cold streams will radiate \lya emission powered by gravitational energy and produce an extended \lya nebula around galaxies without broadening the continuum emission profile. \citet{R12} indicated that the size of \lya nebula depend on the dark matter halo mass $M_{DH}$. In their simulations, the massive ($M_{DH} \geq 10^{12}M_{\odot}$) galaxies have large \lya nebulae ($\gtrsim 100$ kpc) while the less-massive ($M_{DH} \sim 10^{11}M_{\odot}$) galaxies have small \lya nebulae size($\sim 20$ kpc). These cooling radiation dominated models produce relative larger LAHs which are beyond our effective data range. Our work can not test this scenario. Also, in the \citet{Lake15} simulations, there is a small but non-negligible portion of flux from cooling radiation. Our result that UV bright objects have larger halo are consistent with the cold streams origin, but it is also consistent with the center galaxy origin.
The profile of the cooling radiation component is similar to that of the scattered light, we cannot separate these two components through model fitting. One method to estimate the cold stream contribution is to measure the \lya EW at large radii. If cold streams are responsible for the LAHs, the \lya EW values at large radii should be larger than 240 \AA, which is thought to be the maximum value for \lya photons originating from Population II star formation \citep{Malhotra.Rhoads2002}. Our imaging data are not sufficient to draw a conclusion.

Our results agree well with the scatter only model of  \citet{Z11} within their one-halo term effective range.
We do not support the nearby satellite galaxies origin because of the lack of halo detection in broadband ($0 \approx r_{g} << r_{h}$). Due to the relative small fitting range, we can not test the gravitational cooling radiation origin and possible satellite galaxies contribution at large radii. Also, we can not rule out the cooling radiation contribution in the inner region.

\section{Summary}
\label{sec:summary}

We have reported the detection of diffuse LAHs around star-forming galaxies at $z \sim 5.7$ by stacking 310 spectroscopically confirmed LAEs. Our images were taken with Subaru Suprime-Cam in several well studied deep fields. We select LAE candidates using the narrowband technique and carry out follow-up spectroscopic observations using the Magellan/M2FS instrument. Our galaxies have been visually inspected. We also perform detailed simulations and analyses to ensure that the LAHs are not caused by any errors.

We extract the 1D radial profiles of the stacked \lya images and model them using a single-exponential model and a double-exponential model. Our full sample exhibit a LAH scale length of $\sim 2.59$ kpc for the single exponential model and $\sim 2.02$ kpc for the double exponential model. Based on $\chi^2$ test results, All subsamples report LAHs with scale lengths in range of $1.2 \sim 5.3$ kpc except for three subsamples with almost all objects z band undetected. The sizes are relative smaller than those in previous stacking studies, but are consistent with individual LAE measurements from VLT/MUSE. We also find that LAEs with smaller EW, larger \lya luminosity, or larger continuum luminosity tend to have larger halo sizes.

Our results are consistent with some simulation studies that support the scenario in which LAHs originate from scattered light of the central galaxies by H I gas in the CGM. Our LAHs are not likely caused by satellite galaxies because of the lack of halo detection in the continuum band. But we are not able to rule out a possible contribution from satellite galaxies at large radii. Due to a relatively small fitting range, we can not test the gravitational cooling radiation origin.

\appendix

\section{The 1D model}
\label{appendixA}

In this section, we describe the details of our 1D model. We will derive the equations of the 1D model, and address how we produce a continuum-free \lya component and a line-free continuum component in the model. We will explain the reason why we do not subtract continuum flux from the narrowband images, and demonstrate the advantages of our method.

We model an LAE with two components: a \lya emission line and a power-law continuum with slope $\beta = -2$. The flux density of an object 'obj' in the band 'band' is 
$$F_{band}^{obj} = \frac{\int {f_{obj}(\nu)R_{band}(\nu)\, d{\nu}}}{\int{R_{band}(\nu)\, d{\nu}}},$$
where $R_{band}$ is the filter response curve. The model narrowband and broadband flux density $F_{n}$ and $F_{z}$ are
$$F_{n} = F^{\mathrm{Ly}\alpha}_{n} +F^{con}_{n}, $$ and 
$$F_{z} = F^{\mathrm{Ly}\alpha}_{z} +F^{con}_{z}, $$
where $F^{\mathrm{Ly}\alpha}_{n}$ is the \lya line flux in the narrow band, $F^{\mathrm{Ly}\alpha}_{z}$ is the \lya line flux in the broad band, $F^{con}_{n}$ is the continuum flux in the narrow band, and $F^{con}_{z}$ is the continuum flux in the broad band. All these are model data. For real image data, they are denoted as $F_{n(data)}$, $F_{z(data)}$, etc.

We have called  $F^{\mathrm{Ly}\alpha}_{n}$ the continuum-free narrowband flux and $F^{con}_{z}$ the line-free continuum flux. Given the redshift of \lya and the filter response curve , we can calculate the flux ratio of different components in two bands. Both of them are small values,
$$R^{\mathrm{Ly}\alpha}_{n\Rightarrow z} = \frac{F^{\mathrm{Ly}\alpha}_{z}}{F^{\mathrm{Ly}\alpha}_{n}}$$
$$R^{con}_{z\Rightarrow n} = \frac{F^{con}_{n}}{F^{con}_{z}}$$

Therefore,

\begin{eqnarray*}
	F_{n} &=& F^{\mathrm{Ly}\alpha}_{n} +F^{con}_{n} \\
	&=& F^{\mathrm{Ly}\alpha}_{n} + R^{con}_{z\Rightarrow n} * F^{con}_{z} \\
	&=& F^{\mathrm{Ly}\alpha}_{n} + R^{con}_{z\Rightarrow n} * (F_{z} - F^{\mathrm{Ly}\alpha}_{z}) \\ 
	&=& F^{\mathrm{Ly}\alpha}_{n} + R^{con}_{z\Rightarrow n} * F_{z} - R^{con}_{z\Rightarrow n} * F^{\mathrm{Ly}\alpha}_{z}
\end{eqnarray*}
and similarly,		
$$F_{z} = F^{con}_{z} + R^{\mathrm{Ly}\alpha}_{n\Rightarrow z} * F_{n} - R^{\mathrm{Ly}\alpha}_{n\Rightarrow z} * F^{con}_{n}$$

The last terms in the two equations can be ignored because they are the second order, small values compared to the total flux, as shown below,
$$
\lvert \frac{\delta F_{n}}{F_{n}} \rvert
= \frac{R^{con}_{z\Rightarrow n} * F^{\mathrm{Ly}\alpha}_{z}}{F_{n}}
= \frac{R^{con}_{z\Rightarrow n} * F^{\mathrm{Ly}\alpha}_{z}}{F^{\mathrm{Ly}\alpha}_{n} + F^{con}_{n}}
< \frac{R^{con}_{z\Rightarrow n} * F^{\mathrm{Ly}\alpha}_{z}}{F^{\mathrm{Ly}\alpha}_{n}} 
= R^{con}_{z\Rightarrow n} * R^{\mathrm{Ly}\alpha}_{n\Rightarrow z} =  \lvert \frac{\delta F_{z}}{F_{z}} \rvert
$$

This reduces the equations to
\begin{eqnarray}
	F_{n} &=& F^{\mathrm{Ly}\alpha}_{n} + R^{con}_{z\Rightarrow n} * F_{z} \nonumber \\
	F_{z} &=& F^{con}_{z} + R^{\mathrm{Ly}\alpha}_{n\Rightarrow z} * F_{n} \label{eq:app0} 
\end{eqnarray}

In the paper, we use the narrowband NB816, which is outside of the broadband band $z$, so $F^{\mathrm{Ly}\alpha}_{z}$ and $R^{\mathrm{Ly}\alpha}_{n\Rightarrow z}$ are zero. We separate the continuum-free \lya component and line-free continuum component by,

\begin{eqnarray}
	F_{n} &=& F^{\mathrm{Ly}\alpha}_{n} + R^{con}_{z\Rightarrow n} * F_{z} \nonumber \\
	F_{z} &=& F^{con}_{z} \label{eq:app1}
\end{eqnarray}

We then model the continuum-free narrowband flux $F^{\mathrm{Ly}\alpha}_{n}$ using two different models used in previous work.

\begin{itemize}
	\item single exponential model:
		\begin{eqnarray*}
		F^{\mathrm{Ly}\alpha}_{n} &=& S_h \mathrm{exp}(-\frac{r}{r_h}) \\
		F^{con}_{z} &=& S_{z} \mathrm{exp} (-\frac{r}{r_g})
		\end{eqnarray*}
	\item double exponential model:
		\begin{eqnarray*}	
		F^{\mathrm{Ly}\alpha}_{n} &=&  \mathrm{PSF} *(S_h \mathrm{exp}(-\frac{r}{r_h}) + S_g \exp (-\frac{r}{r_g})) \\
		F^{con}_{z} &=& \mathrm{PSF} * S_{z} \mathrm{exp} (-\frac{r}{r_g})
		\end{eqnarray*}
\end{itemize}
Substituting them into \autoref{eq:app1}, we finally get Equations 1 and 2 for MCMC fitting.

In previous work, $F^{\mathrm{Ly}\alpha}_{n}$ is usually obtained by an image subtraction $F^{\mathrm{Ly}\alpha}_{n} = F_{n} - R^{con}_{z\Rightarrow n} * F_{z(data)}$. If our purpose is to fit the confidence interval for the model parameters, this step would be unnecessary and introduce extra statistic error for $F^{\mathrm{Ly}\alpha}_{n}$. It is better to fit the data with a full model instead of separating the data and fitting them with different model components. Therefore, we choose to add the extra flux $R^{con}_{z\Rightarrow n} * F_{z}$(not $R^{con}_{z\Rightarrow n} * F_{z(data)}$) into the model and do MCMC fitting using the full model $F_{n}$ directly.

\begin{figure*}
	\begin{minipage}{0.5\textwidth}
		\includegraphics[keepaspectratio,width=1.0\textwidth]{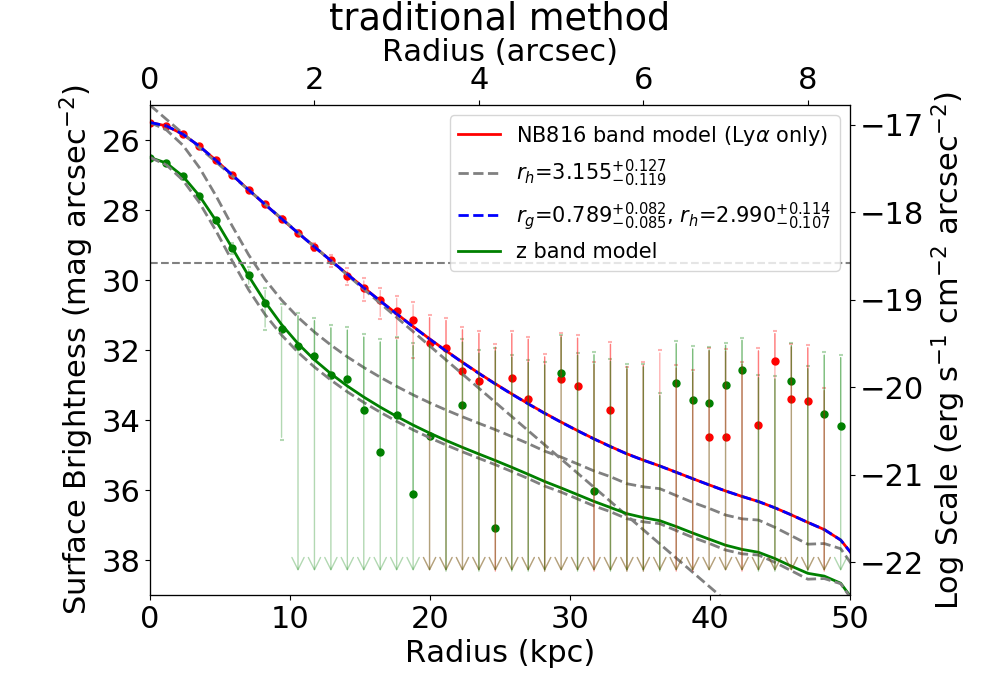}
	\end{minipage}\hfill
	\begin{minipage}{0.5\textwidth}
		\includegraphics[keepaspectratio,width=1.0\textwidth]{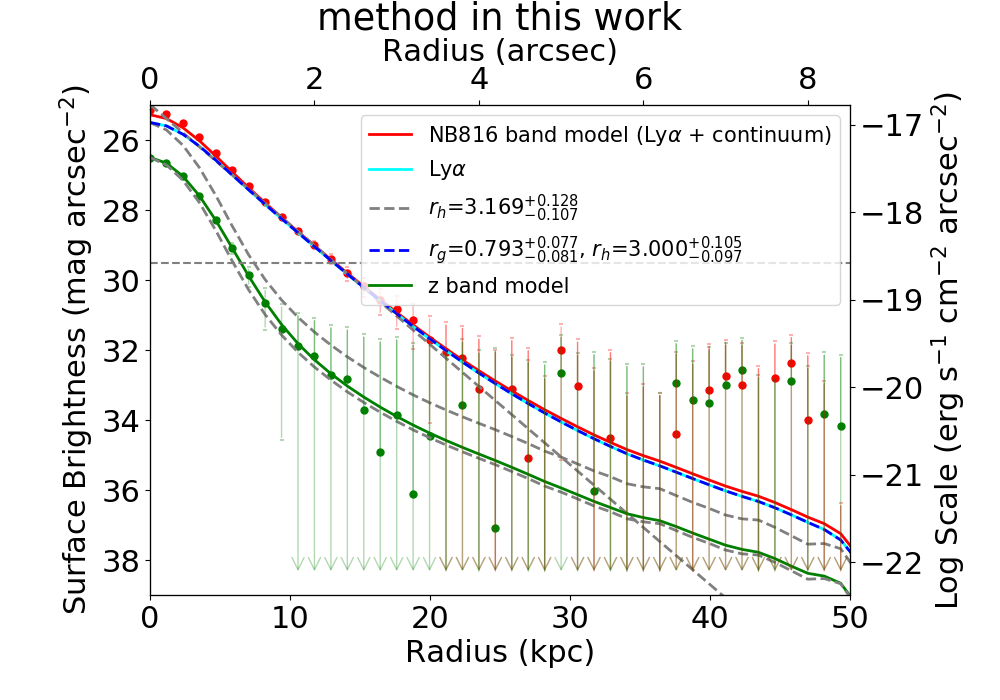}
	\end{minipage}
	\caption{
	  Radial profile and fitting results of one simulation. Left panel: traditional method. Right panel: method in this work. 
	  The red and green dots are the stacked NB816 band and $z$-band radial profiles. The red and green solid lines are the theoretical profiles. The cyan sold line is the \lya component in NB816 band. The gray dashed straight line is the fitting result of the single exponential model. The blue dashed curve is the fitting result of the double exponential model. The horizontal dashed line shows the $1\sigma$ detection limit of the stacked image (per pixel). Compared to our method, the narrowband error from the traditional method is $\sim 15\%$ larger and the confidence interval of the fitting scale length is $\sim 9\%$ larger.
	}
	\label{fig:last}
\end{figure*}

We use mock galaxies to compare our method with the traditional method. The input theoretical profile has an LAH size similar to the typical size in our z\_b subsample ($r_g=0.8$ kpc, $r_h=3.0$ kpc). We randomly select blank regions in our images, mask out objects, and cut them into stamp images. We then convert the 1D theoretical profiles in the NB band and $z$ band to 2D images. The profiles are scaled to match the color of the LAEs in our z\_b subsample. Finally, we add the $z$-band contribution to the NB band according to our LAE spectral model, and put them onto the random background stamps to simulate the errors of the real data. We process 300 mock images using our method and the traditional method. We get consistent and unbiased results from both methods. But compared to our method (See \autoref{fig:last}), the narrowband error from the traditional method is $\sim 15\%$ larger and its confidence interval of the fitting scale length is $5\% \sim 10\%$ larger. Therefore, our method can slightly improve the fitting results.

Also, our model can be applied to any broadband-narrowband combinations. We use \autoref{eq:app1} in this work because NB816 and $z$ are separated, but it is different for other broadband-narrowband combinations. For example, the narrow band NB912 is often used to select $z\sim6.5$ LAEs, and it is located in the center of $z$. The more general equation is \autoref{eq:app0}, in which $F_{n}$ and $F_{z}$ are coupled with each other, and we cannot separate them by a simple subtraction. In this situation, we can do the MCMC fitting iteratively. We first fit $F_{z}$ with the image data $F_{n(data)}$ as the initial value of model $F_{n}$, substitute $F_{z}$ into $F_{n}$ and do another fitting, then substitute $F_{n}$ back into $F_{z}$, and so on. We repeat these steps until they converge. Therefore, our method is more general and can be used for future work with different bands.

In this work, we mainly analyse the results from the weighted-mean stacking algorithm. It is a linear algorithm and thus we should get the same results regardless the fact that the continuum is ``subtracted'' in the 2D processing step or 1D processing step. The main difference is that we calculate the extra flux based on the (smooth) model flux $R^{con}_{z\Rightarrow n} * F_{z}$, while the image subtraction method uses the observed flux $R^{con}_{z\Rightarrow n} * F_{z(data)}$ that contains errors. The theoretical difference is $R^{con}_{z\Rightarrow n} * (F_{z}-F_{z(data)})$, where $(F_{z}-F_{z(data)})$ is the fitting residual of $F_{z(data)}$ and should be a random variable with a mean value of zero. Its error is the error of $F_{z(data)}$.
 

\acknowledgments

We acknowledge support from the National Key R\&D Program of China (2016YFA0400703), the National Science Foundation of China (11533001, 11721303, 11890693), and the Chinese Academy of Sciences (CAS) through a China-Chile Joint Research Fund \#1503 administered by the CAS South America Center for Astronomy in Santiago, Chile. This paper includes data gathered with the 6.5 meter Magellan Telescopes located at Las Campanas Observatory, Chile.
We thank Zheng Zheng for useful discussions.
This research includes data obtained through the Telescope Access Program (TAP), which has been funded by the National Astronomical Observatories of China (the Strategic Priority Research Program `The Emergence of Cosmological Structures', Grant No. XDB09000000), and the Special Fund for Astronomy from the Ministry of Finance.

\bibliography{./references}
\end{document}